\documentclass[aps,prl,twocolumn,superscriptaddress,showpacs,floatfix]{revtex4-1}
\usepackage{makecell}
\usepackage{graphicx}
\usepackage{dcolumn}
\usepackage{bm}
\usepackage{times}
\usepackage{color}
\begin{document}
\bibliographystyle{apsrev4-1}

\title{Quantum oscillations and electronic structures in large Chern number semimetal RhSn}
\author{Sheng Xu}\thanks{These authors contributed equally to this paper}
\affiliation{Department of Physics, Renmin University of China, Beijing 100872, P. R. China}
\affiliation{Beijing Key Laboratory of Opto-electronic Functional Materials $\&$ Micro-nano Devices, Renmin University of China, Beijing 100872, P. R. China}
\author{Liqin Zhou}\thanks{These authors contributed equally to this paper}
\affiliation{Beijing National Laboratory for Condensed Matter Physics and Institute of Physics, Chinese Academy of Sciences, Beijing 100190, P. R. China}
\affiliation{CAS Center for Excellence in Topological Quantum Computation, University of Chinese Academy of Sciences, Beijing 100049, P. R. China}
\author{Huan Wang}
\affiliation{Department of Physics, Renmin University of China, Beijing 100872, P. R. China}
\affiliation{Beijing Key Laboratory of Opto-electronic Functional Materials $\&$ Micro-nano Devices, Renmin University of China, Beijing 100872, P. R. China}
\author{Xiao-Yan Wang}
\affiliation{Department of Physics, Renmin University of China, Beijing 100872, P. R. China}
\affiliation{Beijing Key Laboratory of Opto-electronic Functional Materials $\&$ Micro-nano Devices, Renmin University of China, Beijing 100872, P. R. China}
\author{Yuan Su}
\affiliation{Department of Physics, Renmin University of China, Beijing 100872, P. R. China}
\affiliation{Beijing Key Laboratory of Opto-electronic Functional Materials $\&$ Micro-nano Devices, Renmin University of China, Beijing 100872, P. R. China}
\author{Peng Cheng}
\affiliation{Department of Physics, Renmin University of China, Beijing 100872, P. R. China}
\affiliation{Beijing Key Laboratory of Opto-electronic Functional Materials $\&$ Micro-nano Devices, Renmin University of China, Beijing 100872, P. R. China}
\author{Hongming Weng}\email{hmweng@iphy.ac.cn}
\affiliation{Beijing National Laboratory for Condensed Matter Physics and Institute of Physics, Chinese Academy of Sciences, Beijing 100190, P. R. China}
\affiliation{CAS Center for Excellence in Topological Quantum Computation, University of Chinese Academy of Sciences, Beijing 100049, P. R. China}
\affiliation{Songshan Lake Materials Laboratory, Dongguan, Guangdong 523808, P. R. China}
\author{Tian-Long Xia}\email{tlxia@ruc.edu.cn}
\affiliation{Department of Physics, Renmin University of China, Beijing 100872, P. R. China}
\affiliation{Beijing Key Laboratory of Opto-electronic Functional Materials $\&$ Micro-nano Devices, Renmin University of China, Beijing 100872, P. R. China}


\date{\today}
\begin{abstract}
We report the magnetoresistance, Hall effect, de Haas-van Alphen (dHvA) oscillations and the electronic structures of single crystal RhSn, which is a typical  material of CoSi family holding a large Chern number.  The large unsaturated magnetoresistance is observed with B//[001]. The Hall resistivity curve  indicates that RhSn is a multi-band system with high mobility. Evident quantum oscillations have been observed, from which the light effective masses are extracted. Ten fundamental frequencies are extracted after the fast Fourier transform analysis of the dHvA oscillations with B//[001] configuration. The two low frequencies F$_1$ and F$_2$ do not change obviously and the two high frequencies F$_9$ and F$_{10}$ evolve into four when B rotates from B//[001] to B//[110], which is consistent with the band structure in the first-principles calculations with spin-orbit coupling (SOC). The extracted Berry phases of the relative pockets show a good agreement with the Chern number $\pm4$ (with SOC) in the first-principles calculations. Above all, our studies indicate that RhSn is an ideal platform to study the unconventional chiral fermions and the surface states.

\end{abstract}
\pacs{75.47.-m, 71.30.+h, 72.15.Eb}
\maketitle
\setlength{\parindent}{1em}
\section{Introduction}

Topological semimetals (TSMs) have attracted tremendous attention due to their novel properties in condensed matter physics and materials science\cite{wehling2014dirac,neupane2014observationCd3As2,YLChen2014stableCd3As2,PhysRevLett.113.027603,liang2015ultrahigh,li2015giant,li2015negative,wang2012dirac,liu2014discovery,xiong2015evidence,xiong2016anomalous,wan2011topological,fang2003anomalous,weng2015weyl,xu2015discovery,huang2015weyl,lv2015experimental,ISI:000360709200013,
xu2015discovery2,xu2015experimental,xu2016observation,liu2016evolution,huang2015observation,zhang2016signatures,arnold2016negative}, such as high mobility, nontrivial Berry phase, large magnetoresistance (MR) and/or negative longitudinal MR. In addition to the well-known spin-1/2 Weyl fermion in TaAs family \cite{weng2015weyl,xu2015discovery,huang2015weyl,lv2015experimental,ISI:000360709200013,xu2015discovery2,xu2015experimental,xu2016observation,liu2016evolution,huang2015observation,zhang2016signatures,arnold2016negative}, some new types of fermion with large Chern number, which are spin-1 chiral fermion\cite{manes2012existence,tang2017multiple}, double Weyl fermion\cite{tang2017multiple,xu2016type} and spin-3/2 Rartia-Schwinger-Weyl (RSW) fermion\cite{rarita1941theory,liang2016semimetal,ezawa2016pseudospin}, have been put forward in condensed matter physics.  CoSi and its family materials are predicted to possess these new classes of chiral fermion and later confirmed in experiments\cite{bradlyn2016b,tang2017multiple,pshenay2018band,chang2017unconventional,sanchez2019discovery,rao2019new,schroter2019chiral}.  The SOC induced band splitting is so faint in CoSi that it has not been observed in the angle-resolved photomission spectroscopy (ARPES)\cite{sanchez2019discovery,rao2019new,takane2019observation} or the quantum oscillation measurements\cite{xu2019crystal,wu2019single}.

Motivated by these previous results and discussions, we grew the high quality RhSn single crystals with stronger SOC and studied its transport properties in details. RhSn has the similar bulk band structure with CoSi according to the first-principles calculations. While its Chern number of the chiral fermion at $\Gamma$ and R points is reversed due to the chiral lattice\cite{RhSn}. RhSn holds spin-1 excitation fermion with the Chern number -2 at $\Gamma$ and double Weyl fermion with the Chern number +2 at R in the first Brillouin zone. While the double-degeneracy band splits when the SOC is considered, which makes the spin-1 excitation fermion evolve into a spin-3/2 RSW fermion with Chern number -4 and the double Weyl fermion evolve into a time-reversal (TR) doubling of spin-1 excitation fermion with total Chern number +4, respectively. RhSn  displays a large unsaturated longitudinal magnetoresistance at 2 K and 14 T with B//[001] configuration, which decreases with the B rotating from B$\perp$I to B//I. The nonlinear Hall resistivity and the corresponding analysis indicate that RhSn is a multiband system with high mobility at low temperature. Evident de Haas-van Alphen oscillations and weak Shubnikov-de Haas oscillations (SdH) have been observed. Ten fundamental frequencies have been extracted after the fast Fourier transform (FFT) analysis with B//[001] configuration. The two low frequencies F$_1$ and F$_2$ do not change obviously while the two high frequencies F$_9$ and F$_{10}$ evolve into four high frequencies when B rotates from B//[001] to B//[110]. Thus, the frequencies F$_1$, F$_2$ and F$_9$, F$_{10}$ originate from the electron-like pockets at $\Gamma$ and R, respectively, according to the first-principles calculations. The light cyclotron effective masses are extracted from the fitting by the thermal damping term of the Lifshitz-Kosevich (LK) formula, indicating the possible existence of massless quasiparticles. The extracted Berry phases of the relative F$_1$, F$_9$ and F$_{10}$ pockets show a good agreement with the Chern number $\pm4$ (with SOC) in the first-principles calculations.

\section{Methods and crystal structure}
The high quality single crystals of RhSn were grown by the
Bi flux method. The Rhodium powder, stannum and bismuth granules were put into
the crucible and sealed into a quartz tube with the ratio of 
Rh:Sn:Bi=1:1:16. The quartz tube was heated to 1100 $^{\circ}$C at 60 $^{\circ}$C/h and held for 20 h, then
cooled to 400$^{\circ}$C at 1 $^{\circ}$C/h. The flux was removed by centrifugation. The atomic
composition of RhSn single crystal was checked to be
Rh$:$Sn=1$:$1 by energy dispersive x-ray spectroscopy (EDS, Oxford
X-Max 50). The single-crystal x-ray diffraction (XRD) patterns and
powder XRD patterns were collected from a
Bruker D8 Advance x-ray diffractometer using Cu $K_\alpha$
radiation. TOPAS-4.2 was employed for the refinement. 
The measurements of resistivity and magnetic properties were performed
on a Quantum Design physical property measurement system (QD
PPMS-14T).  The resistivity and Hall measurements were performed using a four-probe method on a long flake crystal which was cut and polished from the crystals as grown. The electrode was made by platinum wire with silver epoxy. The Hall resistivity was obtained from the difference of transverse resistance measurement at the positive and negative fields in order to effectively remove the longitudinal resistivity contribution from the voltage probe misalignment. The first-principles calculations of electronic structure of bulk RhSn were performed by using the Vienna ab initio simulation package (VASP)\cite{kresse1996efficient}, with the generalized gradient approximation (GGA) in the Perdew-Burke-Ernzerhof (PBE) type as the exchange-correlation energy\cite{perdew1996generalized}. The cutoff energy was set to 450 eV and Gamma-centered 10 $\times$ 10 $\times$ 10 $k$ mesh were sampled over the Brillouin zone (BZ) integration. The lattice constant a=5.134${\AA}$ was used for all the calculations and spin-orbit coupling was taken into account. The tight-binding model of RhSn was constructed by the Wannier90 with 4d orbitals of Rh and 5p orbitals of Sn, which was based on the maximally-localized Wannier functions (MLWF)\cite{mostofi2014updated}. The three-dimensional (3D) fermi surface (FS) and FS planes of RhSn were calculated by the WannierTools package\cite{wu2018wanniertools}.

The crystal structure of RhSn is illustrated in Fig. 1(a), which crystallizes in a simple cubic structure with the space group of $P2_13$ (No. 198) \cite{RhSn}. Figure 1(b) shows the single crystal XRD patterns with \{001\}, \{011\}, and \{111\} reflections. The inset of Fig. 1(b) exhibits a picture of the typical grown RhSn crystal with \{001\}, \{011\}, and \{111\} crystal faces, which are highlighted with red dotted line. The powder XRD pattern, as shown in Fig. 1(c), can be well indexed to the structure of RhSn with space group $P2_13$, and the refined lattice parameter is a=b=c=5.13{\AA}.

\section{Results and Discussions}

As displayed in Fig. 2(a), the temperature-dependent resistivity of RhSn exhibits metallic behavior with large residual resistivity ratio (RRR$\approx$24), indicating the quality of the grown crystal.  Figure 2(b) presents the MR=($\rho_{xx}(B)-\rho_{xx}(0))/\rho_{xx}(0)$ versus magnetic field at various temperatures. The MR reaches 450\% at 2 K and 14 T with B//c configuration. The SdH oscillations are observed at 2 K and high magnetic field. The magnetic field-dependent MR with the magnetic field tilted from B$\bot$I ($\theta=0^o$) to B//I ($\theta=90^o$) is exhibited in Fig. 2(c). The MR decreases with the increasing $\theta$ and is maximized (minimized) for B$\bot$I (B//I), respectively, while the negative longitudinal MR with B//I configuration has not been observed due to the trivial hole-like pockets induced orbit MR at M and near $\Gamma$. In order to thoroughly identify the characteristics of the carriers in RhSn, we performed temperature-dependent Hall resistivity measurements. The Hall resistivity $\rho_{xy}$ curves of RhSn at various temperatures from 2 K to 150 K are shown in Fig. 2(d). Further, the deviation from the linear behavior of the Hall resistivity indicates that RhSn is a multi-band system. Thus, the two-band model is used to describe the Hall conductivity,

\begin{figure}
\centering
  \includegraphics[width=0.48\textwidth]{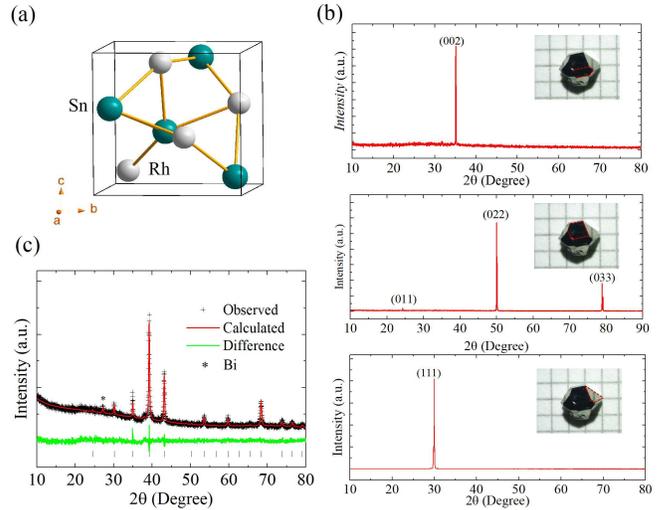}\\
  \caption{(a) Crystal structure of RhSn. (b) Single crystal XRD pattern of RhSn with different crystal faces. Insets show the picture of the grown crystal. (c)
Powder XRD pattern and the Rietveld refinement of RhSn. The
value of R$_{wp}$ is 5.667\%. }\label{1}
\end{figure}

\begin{figure}[htbp]
\centering
  \includegraphics[width=0.48\textwidth]{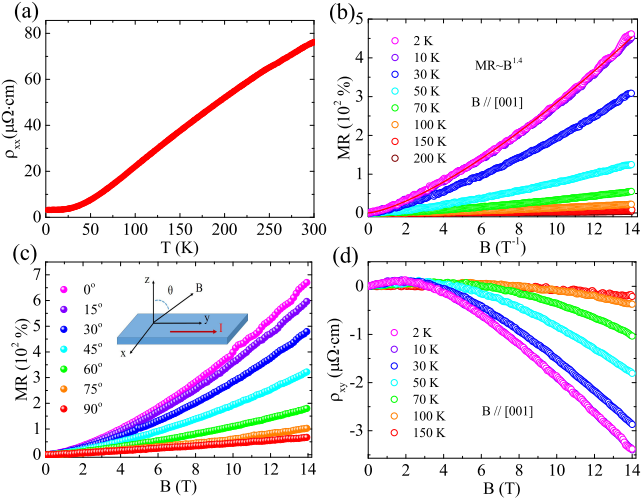}\\
\caption{(a) Temperature dependence of the resistivity $\rho_{xx}$. (b) Magnetic field dependence of MR at different temperatures.
(c) Magnetic field-dependent MR at 2 K with magnetic field titled from B$\bot$I ($\theta=0^o$) to B//I ($\theta=90^o$). Inset shows the definition of $\theta$. (d) Magnetic field-dependent Hall resistivity at various temperatures.}\label{2}
\end{figure}

\begin{equation}\label{equ3}
\sigma_{xy}=(\frac{n_h \mu_h^2}{1+(\mu_h B)^2}-\frac{n_e \mu_e^2}{1+(\mu_e B)^2})eB
\end{equation}
where $n_{e,h}$ and $\mu_{e,h}$ represent the concentration and
mobility of electrons and holes, respectively. As shown in the Fig. 3(a), the fitted red curves are consistent with the experimental dots, from which the temperature-dependent concentrations and mobility are extracted and plotted in Figs. 3(b) and 3(c). The concentration and mobility of the two types of carriers increase with the decreasing temperature. At 2 K, $n_e=4.6\times10^{20} cm^{-3}$ which is almost two times larger than $n_h=2.2\times10^{20} cm^{-3}$, and the $\mu_e$ and
$\mu_h$ are 1895 $cm^2V^{-1}s^{-1}$ and 3609 $cm^2V^{-1}s^{-1}$, respectively.
\begin{figure}[htbp]
\centering
  \includegraphics[width=0.48\textwidth]{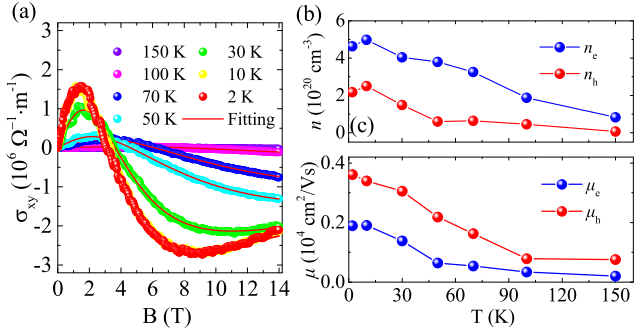}\\
\caption{(a) Field dependence of the Hall conductivity $(\sigma_{xy}=\rho_{xy}/((\rho_{xy})^2+(\rho_{xx})^2))$ at various temperatures. (b) and (c) The temperature dependence of carrier densities and mobility of the electrons and holes extracted from the two-band model fitting.}\label{2}
\end{figure}

\begin{figure}[htbp]
\centering
  \includegraphics[width=0.48\textwidth]{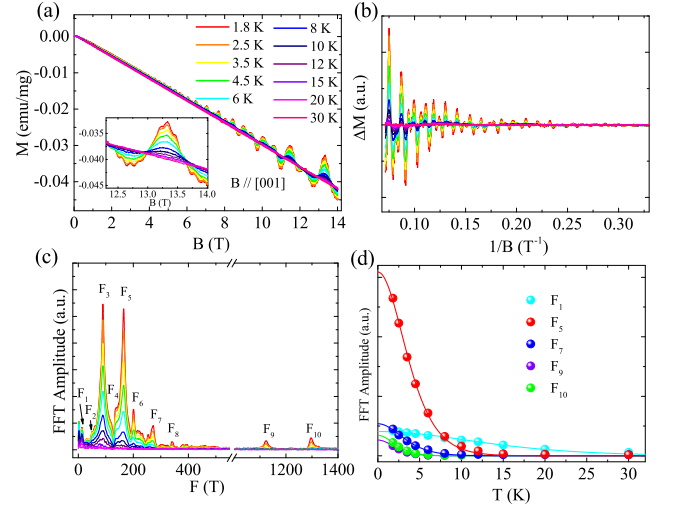}\\
\caption{(a) The dHvA oscillations at various temperatures with
\emph{B}//[001] configuration. (b) The amplitudes of dHvA oscillations as a function of 1/B. (c) The FFT spectra
of the oscillations. (d) The temperature dependence of relative FFT amplitude of the frequencies.}\label{2}
\end{figure}

\begin{table*}
  \centering
\caption{The parameters extracted from dHvA oscillations in RhSn with B//[001] configuration
. F is the frequency of dHvA oscillations; $m^*$/$m_e$ is the ratio
of the effective mass to the electron mass; $\phi$$_B$ is the Berry phase. }
  \label{oscillations}
  \setlength{\tabcolsep}{3.4mm}{
  \begin{tabular}{ccccccccccccc}
  \hline
  \hline
   &                           &$F_1$     & $F_2$    & $F_3$ & $F_4$ & $F_5$ & $F_6$ & $F_7$ & $F_8$ & $F_9$    & $F_{10}$ &\\
   \hline
  &$F(T)$                      &13.1      &49.4      &89.2   &134.8  &164.9  &200.6  &272.2  &340.6  &1123.4    &1296.7&\\
  &$m^*$/$m_e$                 &0.038     &/         &/      &/      &0.136  &/      &0.153  &/      &0.229     &0.224&\\
  &$\phi_{B}$                  &2.11$\pi$ &/         &/      &/      &/      &/      &/      &/      &2.10$\pi$ &2.02$\pi$&\\
  &$A_{F}$ $10^{-3}{\AA}^{-2}$ &1.25      &4.71      &8.51  &12.87   &15.74  &19.15  &25.98  &32.51  &107.24    &123.78&\\
  \hline
  \hline
  \end{tabular}}
  \end{table*}

\begin{figure}
	\centering
	\includegraphics[width=0.48\textwidth]{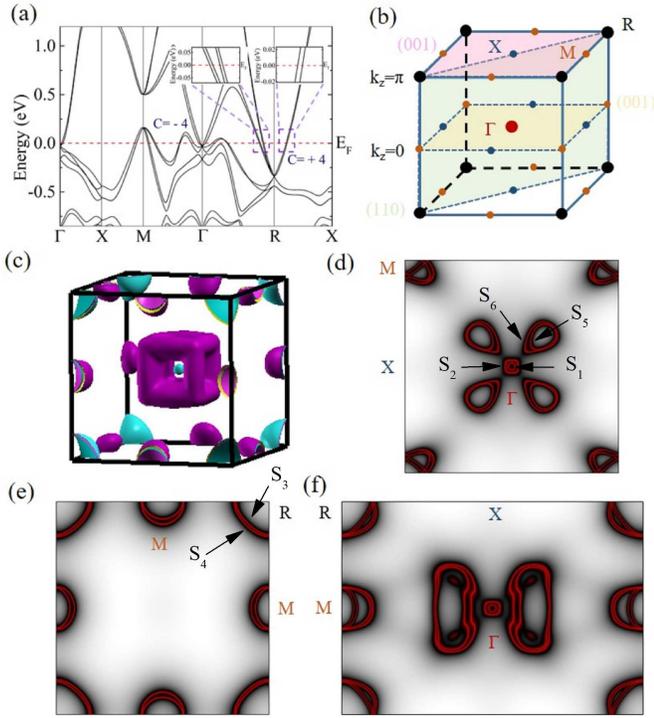}\\
	\caption{(a) Calculated bulk band structure of RhSn along high-symmetry lines with SOC. (b) Brillouin zone of RhSn. (c) Calculated FSs in the bulk Brillouin zone. (d)(e)(f) Calculated FSs in $k_z$=0 plane, $k_z$=$\pi$ plane and $\Gamma$-X-M-R plane, respectively.}\label{5}
\end{figure}

\begin{figure}[htbp]
	\centering
	\includegraphics[width=0.48\textwidth]{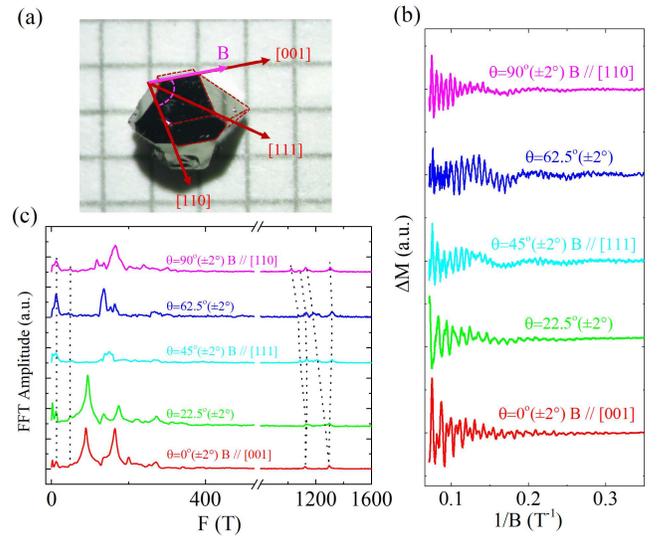}\\
	\caption{The sketch map of the respective rotating B direction along crystallographic direction. (b) Angle-dependent dHvA oscillations at low temperature. (c) Corresponding FFT spectra for B rotating from B//[001] to B//[110]}\label{2}
\end{figure}

\begin{figure}
	\centering
	\includegraphics[width=0.48\textwidth]{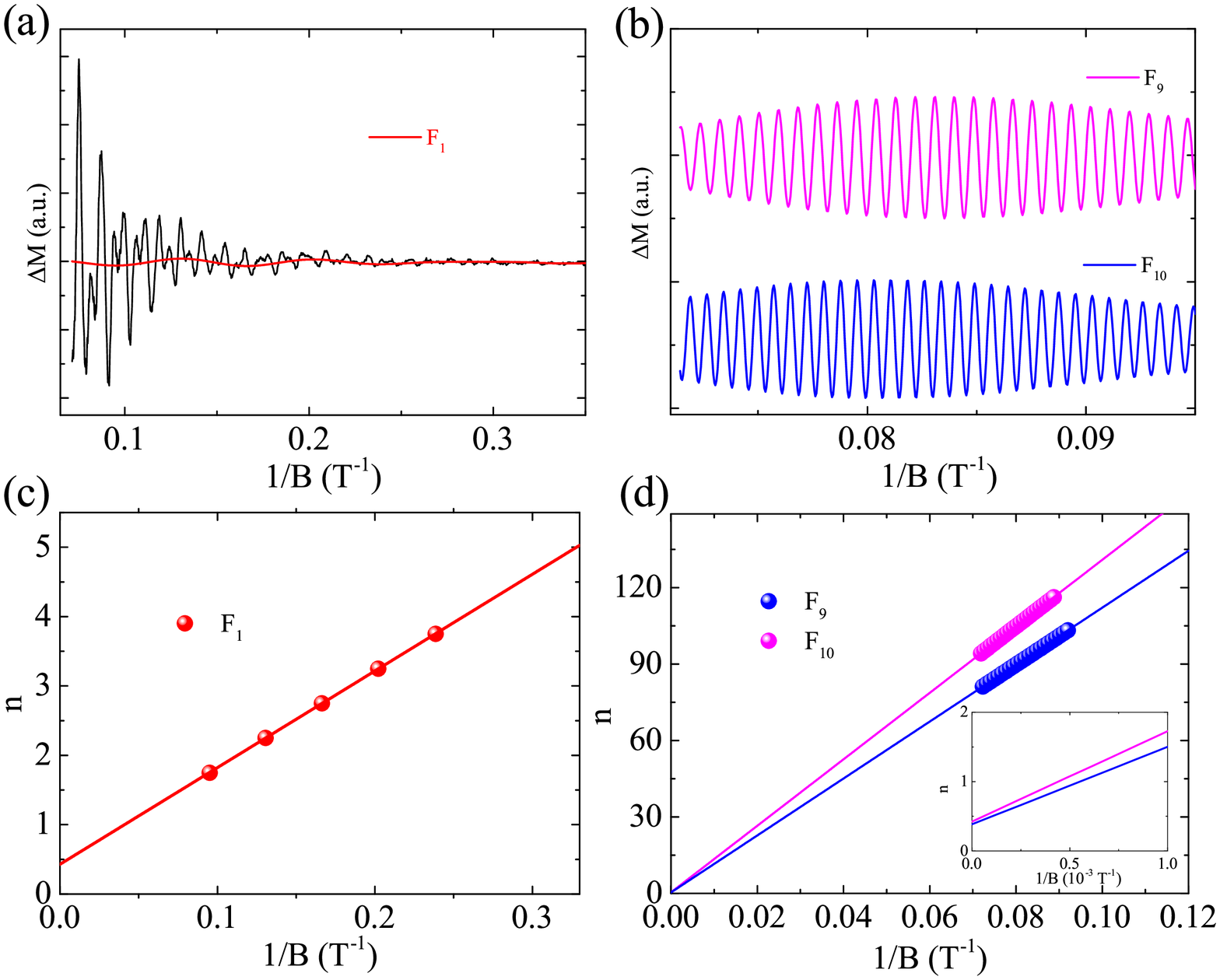}\\
	\caption{(a) The filtered frequencies of $F_2$ (red curve) and the background of the observed  $\Delta$M versus 1/B (black curve) when B//[001]. (b) The filtered frequencies of $F_9$ and $F_{10}$. (c) and (d) The Landau level index fan diagram for the filtered frequencies. }\label{6}
\end{figure}

ARPES is a powerful tool to study the Fermi surface properties. While the SOC induced band splitting has not observed in our former ARPES measurements, because the band splitting is not evident enough and the two splitting bands are beyond the ARPES resolution\cite{RhSn}. Quantum oscillation experiment provides another effective method to study the characteristics of FS, since the oscillation frequency is proportional to the extreme cross section of FS normal to the magnetic field according to the Onsager relation $F=(\phi{_0}/2\pi^2)=(\hbar/2\pi e)A$. Figure 4(a) presents the isothermal magnetization of a typical single crystal with B//c configuration, which exhibits evident dHvA oscillations. The oscillatory components of magnetization are obtained and displayed in Fig. 4(b) by subtracting a smooth background. Ten fundamental frequencies are obtained after the FFT analysis when the field is applied along [001] direction. Thus, the corresponding extreme cross section area of F$_1$, F$_2$, F$_9$ and F$_{10}$ are $1.25\times10^{-3}$ {\AA}$^{-2}$, $4.71\times10^{-3}$ {\AA}$^{-2}$, $107.24\times10^{-3}$ {\AA}$^{-2}$ and $123.78\times10^{-3}$ {\AA}$^{-2}$, respectively. More details are displayed in Table I. There is a very low frequency about 2 T in the FFT spectra which is not the intrinsic dHvA oscillations signal and comes from the data processing. The oscillatory components versus 1/B of dHvA oscillations can be described by the LK formula\cite{shoenberg2009magnetic}:

\begin{equation}\label{equ1}
\Delta M \propto -B^{1/2}\frac{\lambda T}{sinh(\lambda T)}e^{-\lambda T_D}sin[2\pi(\frac{F}{B}-\frac{1}{2}+\beta+\delta)]
\end{equation}
where $\lambda=(2\pi^2 k{_B} m^*)/(\hbar eB)$. $T{_D}$ is the Dingle
temperature. The value of $\delta$ depends on the dimensionality,
$\delta=0$ for the 2D system and $\delta=\pm1/8$ for the 3D system.
$\beta=\phi_B/2\pi$ and $\phi_B$ is the Berry phase. Figure 4(c) shows the temperature-dependent FFT amplitudes
and the fitting by the thermal factor $R_T=(\lambda T)/sinh(\lambda T)$ in LK formula. The effective masses are estimated to
be $m^*_{F_1}=0.038m_e$, $m^*_{F_9}=0.229m_e$ and $m^*_{F_{10}}=0.224m_e$, which are exhibited in Table I.

Figures 5(a) and 5(c) exhibit the calculated band structure and the FSs of RhSn with SOC in Brillouin zone. There are electron-like FSs at $\Gamma$ and R points, and the rest are the hole-like FSs. Figures 5(d) and 5(e) exhibt the calculated FSs in $k_z$=0 and $k_z$=$\pi$ plane. There are two smallest FSs extreme cross sections S$_1$, S$_2$ at $\Gamma$ point, and two largest FSs extreme cross sections S$_3$, S$_4$ at R point, as shown in Figs. 5(d) and 5(e). Their calculated area are about S$_1$=$1.3\times10^{-3}$ {\AA}$^{-2}$, S$_2$=$4.9\times10^{-3}$ {\AA}$^{-2}$, S$_3$=$103.5\times10^{-3}$ {\AA}$^{-2}$ and S$_4$=$118.4\times10^{-3}$ {\AA}$^{-2}$, respectively. It means that when B//[001] the calculated oscillation frequencies of these four extreme cross sections are 13.6 T, 51.3 T, 1084.2 T and 1240.3 T, respectively, according to the Onsager relation, which are comparable to the observed two smallest frequencies F$_1$ (13.1 T), F$_2$ (49.4 T) and two largest frequencies F$_9$ (1123.4 T), F$_{10}$ (1296.7 T). By comparing the results between the calculations and experiments, we conclude that the frequency F$_1$ in dHvA oscillations with B//[001] configuration originates from S$_1$, F$_2$ originates from S$_2$, F$_9$ originates from S$_3$ and F$_{10}$ originates from S$_4$.

In strong magnetic field, the electron can tunnel from an orbit on one part of a Fermi surface to an orbit on another separated by a small energy gap, which  brings into the existence of new orbits. It is considered as magnetic breakdown\cite{shoenberg2009magnetic}. As shown in Fig. 5(d), the gaps between the FSs S$_3$ and S$_4$, teardrop-shaped FSs S$_5$ and S$_6$, or the ellipse FSs extreme cross sections at M point are very small, and the magnetic breakdown is easy to happen. It is hard to exactly verify the observed frequencies from F$_3$ to F$_8$. Since all the frequencies from F$_3$ to F$_8$ originate from the trivial hole-like pockets, the analysis was mainly focused on the frequencies F$_1$, F$_2$, F$_9$ and F$_{10}$ which stem from the topological non-trivial electron-like pockets.

Angle-dependent dHvA oscillations measurements are applied to further study the FS characteristics of RhSn. The crystallographic orientation [110], [111] and [001] in one single crystal sample is determined from the measurements and analysis of single XRD patterns, as shown in Fig. 6(a). The orientation of magnetic field B is always within \{110\} plane and rotates from [001] to [110] in the measurements of angle-dependent dHvA oscillations. $\theta$ is the angle between the magnetic field B and [001].  $\theta=0^{\circ}$ when B//[001], $\theta=45^{\circ}$ when B//[111], and  $\theta=90^{\circ}$ when B//[110]. As exhibited in Figs. 5(d) and 5(f), there are two smallest FSs extreme cross sections (S$_{1}$, S$_{2}$) at $\Gamma$ point with B//[001] and B//[110]. It shows a good agreement with the result of angle-dependent dHvA quantum oscillations as the frequencies F$_1$ (S$_1$) and F$_2$ (S$_2$) are almost unchanged with B rotating from along [001] to [110] within \{110\} plane, as shown in the Fig. 6(c). Figures 5(d) and 5(f) also illustrate that there are two largest FSs extreme cross sections (S$_{3}$, S$_{4}$)  at R point with B//[001] corresponding to the two largest frequencies in quantum oscillations, and four FSs extreme cross sections with B//[110] corresponding to the four frequencies in quantum oscillations, respectively. The quantum oscillation frequencies F$_9$ (S$_3$) and F$_{10}$ (S$_4$)  evolve into four frequencies when B rotates from [001] to [110], which is consistent with the calculations. In conclusion, by combining the analysis of angle-dependent dHvA quantum oscillations and the first-principles calculations, we confirm that the frequencies F$_1$, F$_2$, F$_9$ and F$_{10}$ in the quantum oscillations originate from the FSs at $\Gamma$ and R points. 

Berry phase extracted from quantum oscillations is an important parameter of topological materials in transport measurements. In trivial metals, the Berry phase is 0 or 2$\pi$. In the spin-1/2 Weyl semimetal with Chern number $\pm1$, the Berry phase is $\pi$ like TaAs family materials. RhSn holds TR doubling of spin-1 excitation fermion at R and spin -3/2 RSW fermion at $\Gamma$ with the Chern number $\pm4$ (with SOC), thus the Berry phase extracted from the frequencies F$_1$, F$_2$, F$_9$ and F$_{10}$ should be 4$\pi$ in theory. According to Lifshitz-Onsager quantization rule $n=A_F(\hbar/2\pi eB)-1/2+\beta+\delta$, the Berry phase can be extracted from the intercept of the linear extrapolation. The maximum of $\triangle M$ correspond to the Landau indices of n+1/4, and $\delta=-1/8$ since the F$_1$, F$_9$ and F$_{10}$ bands originate from the electron pockets. The frequencies F$_1$, F$_9$ and F$_{10}$ are filtered from the oscillations and the corresponding Berry phases are extracted by the LL index fan diagram. The intercepts of theses bands are 0.429, 0.423 and 0.384. Thus, the Berry phases of F$_1$, F$_9$ and F$_{10}$ are 2.11$\pi$, 2.10$\pi$ and 2.02$\pi$, respectively, which show a good agreement with the Chern number $\pm4$ (with SOC) in the first-principles calculations. The frequency F$_2$ is close to F$_3$ and its FFT amplitude is greatly affected by F$_3$, so the effective mass and Berry phase analysis are not applied on F$_2$.

\section{Summary}

In conclusion, we have successfully synthesized the high quality single crystals of RhSn which hold a spin-3/2 RSW fermion at $\Gamma$ (Chern number -4) and the TR doubling of spin-1 excitation at R (total Chern number +4) according to the first-principles calculations with SOC. It shows a metallic behavior at zero field. The large unsaturated MR at 2 K and 14 T has been observed. The analysis of Hall effect and dHvA oscillations indicate that RhSn is a multi-band system. SOC induced band splitting is observed by our quantum oscillation measurements. When B rotates from B//[001] to B//[110] within \{110\}, the two lowest frequencies of dHvA oscillations F$_1$ and F$_2$ do not  change obviously while the two highest frequencies F$_9$ and F$_{10}$ evolve into four, which is in agreement with the first-principles calculations. Thus, the frequencies F$_1$ and F$_2$ stem from the electron pockets at $\Gamma$ while F$_9$ and F$_{10}$ originate from the electron pockets at R in the first Brillouin zone. The extracted Berry phases $\sim$ 2$\pi$ are in agreement with the calculated Chern number $\pm$4.

\section{Acknowledgments}

We thank T. Qian and Y. J. Sun for helpful discussions. This work is supported by the Ministry of Science and Technology of China (2019YFA0308602, 2018YFA0305700 and 2016YFA0300600), the National Natural Science Foundation of China (No.11874422, No.11574391), the Fundamental Research Funds for the Central Universities, and the Research Funds of Renmin University of China (No.19XNLG18, No.18XNLG14). H.W. also acknowledges support from the Chinese Academy of Sciences under grant number XDB28000000, the Science Challenge Project (No.  TZ2016004), the K. C. Wong Education Foundation (GJTD-2018-01), Beijing Municipal Science \& Technology Commission (Z181100004218001) and Beijing Natural Science Foundation (Z180008).
\bibliography{Bibtex}
\end{document}